\documentclass{article}
\usepackage{spconf,amsmath,graphicx}

\usepackage{graphicx}
\usepackage{amssymb,amsmath,bm}
\usepackage{textcomp}
\usepackage{booktabs}
\usepackage{subfigure} 

\title{Small-footprint Keyword Spotting with\\ Graph Convolutional Network}
\name{Xi Chen$^{1}$, Shouyi Yin$^1$, Dandan Song$^{2}$, Peng Ouyang$^2$, Leibo Liu$^1$, Shaojun Wei$^1$}
\address{
  $^1$Tsinghua University $^2$TsingMicro Co. Ltd.}
\begin{document}
%
\maketitle
\begin{abstract}
Despite the recent successes of deep neural networks, it remains challenging to achieve high precision keyword spotting task (KWS) on resource-constrained devices. In this study, we propose a novel context-aware and compact architecture for keyword spotting task. Based on residual connection and bottleneck structure, we design a compact and efficient network for KWS task. To leverage the long range dependencies and global context of the convolutional feature maps, the graph convolutional network is introduced to encode the non-local relations. By evaluated on the Google Speech Command Dataset, the proposed method achieves state-of-the-art performance and outperforms the prior works by a large margin with lower computational cost.
\end{abstract}

\section{Introduction}\label{sec:intro}

 With the rapid development of speech recognition in recent years, speech interface on smart devices is becoming increasingly popular, such as Google Search by Voice~\cite{schalkwyk2010your}, intelligent loudspeakers and mobile assistants, etc. These various smart devices enable users to obtain a fully hands-free experience by continuously listening to specific keywords to initiate voice input. \textit{Keyword spotting (KWS)}, or \textit{keyword detection}, is a task that aims to detect pre-defined keywords in a stream of audio. A practical on-device KWS module should be highly accurate, low-latency and small footprint for the deployment, while operating on the low-power and resource-constrained devices.

Conventional study for KWS task focuses on keyword/filter hidden Markov model (HMM)~\cite{rohlicek1989continuous, rose1990hidden, wilpon1991improvements, silaghi1999iterative, silaghi2005spotting}. In these approaches, HMMs are trained for both keyword and non-keyword speech segments, respectively. During the runtime, the Viterbi search algorithm is applied to generate the best path in the decoding graph, which could be computationally expensive due to the complexity of the HMMs topology. Several KWS systems use a large vocabulary continuous speech recognizer (LVCSR) to generate a rich lattices, and search the keyword among all possible paths in the lattices~\cite{miller2007rapid, parlak2008spoken, mamou2007vocabulary, gales2014speech}.

In recent years, deep neural networks (DNNs) have been proven to yield significant improvement in KWS task. In~\cite{chen2014small}, Chen et al. propose a small footprint approach called Deep KWS. Specifically, a DNN-based acoustic model is trained to directly predict the frame-level posteriors of sub-keywords followed by a posterior handling method, which produces a final confidence score. This effective idea outperforms the traditional keyword/filter HMMs and is highly attractive to run on devices with a small footprint and low latency. 

After that, more advanced architectures have been successfully applied to KWS task as an alternative to the DNN-based acoustic model. In~\cite{sainath2015convolutional, zhang2017hello}, convolutional neural networks (CNNs) are utilized under limited memory footprint as well as computational resource scenarios. In~\cite{tang2018deep}, Tang et al. introduce the idea of deep residual network (ResNet) to achieve a trade-off between the model footprint and prediction performance. These methods have demonstrated computational efficiency but failed in capturing local receptive fields and short range context. Various attempts have also been made to build a KWS system with recurrent neural networks (RNNs)~\cite{fernandez2007application, woellmer2013keyword, baljekar2014online, sun2016max, he2017streaming}, which is capable of modeling longer temporal context information. However, RNNs may suffer from state saturation while facing continuous input stream, increasing computational cost and detection latency.

In this work, we aim to address the aforementioned limitations of convolution by incorporating the long-range context information with \textit{graph convolutional network (GCN)~\cite{Kipf2016Semi}}. 
The non-local relations can be encoded through a densely-connected GCN module with attention mechanism, which estimates the feature context by message passing on the underlying graph. Intuitively, for the feature at a certain position, it is updated via aggregating features at all positions with weighted summation, where the weights are decided by the feature similarities between the corresponding two positions. Thus, any two positions with similar features can contribute mutual improvement regardless of their distance in spatial dimension, which is different from the progressive behavior of recurrent and convolutional operations. 

Inspired by the ResNet~\cite{he2016deep}, we propose a compact and efficient convolutional network (denoted as CENet) by utilizing the bottleneck architecture with narrow structure. The bottleneck architecture is proposed in~\cite{he2016deep} to reduce the model complexity by introducing a $1\times 1$ convolutional layer, which is responsible for the reducing and restoring dimensions. Different from the prior works, we also investigate the effects of the narrower structure for small-footprint KWS task. 

\begin{table*}[th]
	\center
	\caption{Configuration of the CENet Baseline. (1,16) means the number of input channel and output channel is 1 and 16.}
	\resizebox{0.9\textwidth}{!}{
	\begin{tabular}{l|c|c|c|c|c}
	\toprule[0.3mm]
	Model & \#Parameters  & Initial & Block-1 & Block-2 & Block-3 \\
	 \midrule
	 CENet-6 & 16.2K & $\begin{bmatrix} 3\times 3,(1,16) \\ 
									 \end{bmatrix}$  & $\begin{bmatrix} 1\times 1,(16,8) \\ 
	 									  3\times 3,(8,8) \\
	 									  1\times 1,(8,16) \\
									 \end{bmatrix} \times 1,
									 \begin{bmatrix}
									 	1\times 1,(16,8) \\ 
	 									  3\times 3,(8,8) \\
	 									  1\times 1,(8,32) \\
									 \end{bmatrix} \times 1$ 
					   & $\begin{bmatrix} 1\times 1,(32,8) \\ 
	 									  3\times 3,(8,8) \\
	 									  1\times 1,(8,32) \\
									 \end{bmatrix} \times 1,
									 \begin{bmatrix}
									 	1\times 1,(32,8) \\ 
	 									  3\times 3,(8,8) \\
	 									  1\times 1,(8,48) \\
									 \end{bmatrix} \times 1$ 
					  & $\begin{bmatrix} 1\times 1,(48,12) \\ 
	 									  3\times 3,(12,12) \\
	 									  1\times 1,(12,48) \\
									 \end{bmatrix} \times 1,
									 \begin{bmatrix}
									 	1\times 1,(48,12) \\ 
	 									  3\times 3,(12,12) \\
	 									  1\times 1,(48,64) \\
									 \end{bmatrix} \times 1$ \\\midrule
	 CENet-24 & 44.3K & $\begin{bmatrix} 3\times 3,(1,16) \\ 
									 \end{bmatrix}$  & $\begin{bmatrix} 1\times 1,(16,8) \\ 
	 									  3\times 3,(8,8) \\
	 									  1\times 1,(8,16) \\
									 \end{bmatrix} \times 7,
									 \begin{bmatrix}
									 	1\times 1,(16,8) \\ 
	 									  3\times 3,(8,8) \\
	 									  1\times 1,(8,32) \\
									 \end{bmatrix} \times 1$ 
					   & $\begin{bmatrix} 1\times 1,(32,8) \\ 
	 									  3\times 3,(8,8) \\
	 									  1\times 1,(8,32) \\
									 \end{bmatrix} \times 7,
									 \begin{bmatrix}
									 	1\times 1,(32,8) \\ 
	 									  3\times 3,(8,8) \\
	 									  1\times 1,(8,48) \\
									 \end{bmatrix} \times 1$ 
					  & $\begin{bmatrix} 1\times 1,(48,12) \\ 
	 									  3\times 3,(12,12) \\
	 									  1\times 1,(12,48) \\
									 \end{bmatrix} \times 7,
									 \begin{bmatrix}
									 	1\times 1,(48,12) \\ 
	 									  3\times 3,(12,12) \\
	 									  1\times 1,(48,64) \\
									 \end{bmatrix} \times 1$ \\\midrule
	CENet-40 & 61K  & $\begin{bmatrix} 3\times 3,(1,16) \\ 
									 \end{bmatrix}$  & $\begin{bmatrix} 1\times 1,(16,8) \\ 
	 									  3\times 3,(8,8) \\
	 									  1\times 1,(8,16) \\
									 \end{bmatrix} \times 15,
									 \begin{bmatrix}
									 	1\times 1,(16,8) \\ 
	 									  3\times 3,(8,8) \\
	 									  1\times 1,(8,32) \\
									 \end{bmatrix} \times 1$ 
					   & $\begin{bmatrix} 1\times 1,(32,8) \\ 
	 									  3\times 3,(8,8) \\
	 									  1\times 1,(8,32) \\
									 \end{bmatrix} \times 15,
									 \begin{bmatrix}
									 	1\times 1,(32,8) \\ 
	 									  3\times 3,(8,8) \\
	 									  1\times 1,(8,48) \\
									 \end{bmatrix} \times 1$ 
					  & $\begin{bmatrix} 1\times 1,(48,12) \\ 
	 									  3\times 3,(12,12) \\
	 									  1\times 1,(12,48) \\
									 \end{bmatrix} \times 7,
									 \begin{bmatrix}
									 	1\times 1,(48,12) \\ 
	 									  3\times 3,(12,12) \\
	 									  1\times 1,(48,64) \\
									 \end{bmatrix} \times 1$ \\
 	\bottomrule[0.3mm]
	\end{tabular}
	\label{tab:model_details}}
\end{table*}

By combining the contextual feature augmentation module GCN with the CENet, a compact but efficient model(denoted as CENet-GCN) is proposed for the KWS. We validate our method on the Google Speech Command dataset~\cite{warden2018speech} with a series of comprehensive experiments. The empirical results and ablative study demonstrate the superior of our method over the prior state-of-the-art approaches with much fewer parameters and simpler network structure. 
The main contributions of this work can be summarized as followed:
\begin{itemize}
	\item We propose a compact and efficient convolutional network for small-footprint KWS by utilizing the bottleneck structure.
	\item We introduce the GCN to capture the long range dependencies and achieve the contextual feature augmentation.
\end{itemize}
\vspace{-3mm}
\section{Method} \label{sec:method}

\noindent We describe our method for building a compact and efficient network to achieve the small footprint KWS system. Firstly, we start with a brief introduction of the KWS task and the preprocessing strategies for audio feature generation in \textbf{Sec.~\ref{subsec:problem}}. Next, we detail our compact and efficient network structure based on residual connection and bottleneck structure in \textbf{Sec.~\ref{subsec:basic}}. Finally, we describe the method of the feature enhancement using GCN in \textbf{Sec.~\ref{subsec:gnn}}. To our knowledge, we are the first to apply the GCN approach for KWS task.

\vspace{-2mm}

\subsection{KWS Task and Preprocessing}\label{subsec:problem}
\indent KWS is a task of detecting the pre-defined keyword in audio stream. The CNN-based KWS task includes two phases: \textit{1) pre-processing phase}, \textit{2) CNN-based classifier.} 

In the pre-processing phase, firstly, a band-pass filter of 20Hz/4kHz is used to reduce noise. After that, the Mel-Frequency Cepstrum Coefficient (MFCC) feature is constructed using a 30ms window size and 10ms frame shift. We denote the input MFCC feature as $\mathbf{I}$ feeding into the neural network. In this work $\mathbf{I} \in \mathbb{R}^{t\times f}$, where $t$ is 101 and $f$ is 40. The $f$ is the dimension of the MFCC feature, and the $t$ is the number of frames.

The KWS task can be formulated as a classification problem:
\begin{align}
	\mathbf{y} = \mathcal{F}(\mathbf{I}, \Theta)
\end{align}
where $\mathbf{y}\in \mathbb{R}^l$ is the prediction of the audio sequence, and $\mathcal{F}$ is the mapping function from the input feature space to the label space, which is implemented with neural network in our method. $l$ is the number of classes to be classified. Our goal is to learn the network parameter $ \Theta$ of the mapping function $\mathcal{F}$.



\begin{figure}[t]
\begin{center}
\includegraphics[width=0.8\linewidth]{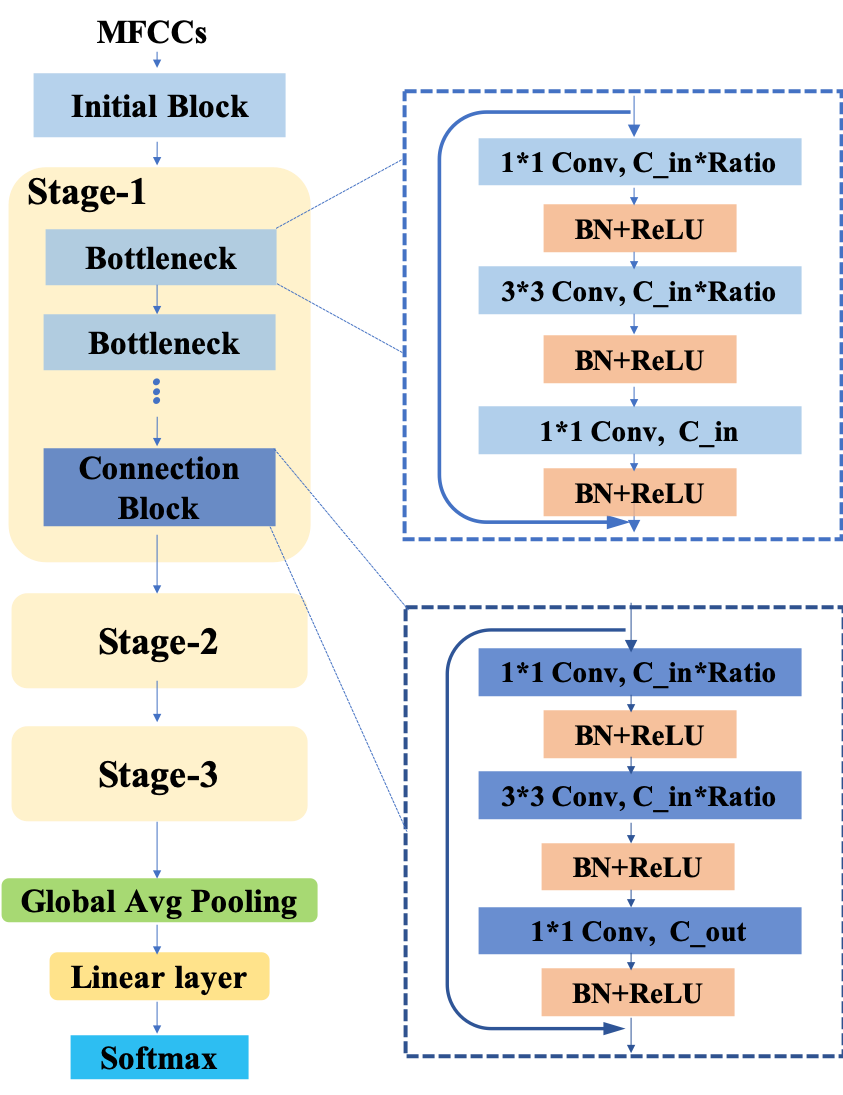}

\end{center}
   \caption{Framework of our CENet.}
\label{fig:short}
\end{figure}


\vspace{-2mm}
\subsection{Compact and Efficient Neural Network}\label{subsec:basic}
Our goal is to learn a KWS classifier with the compact structure and low computational cost. 
Motivated by the prior works~\cite{tang2018deep, he2016deep}, we follow the key idea of the ResNet~\cite{he2016deep}, which postulates that learning residual of the network is easier than the original mapping, to design a family of models for the KWS task in the resource-limited environment. It is worth noting that ResNet~\cite{he2016deep} replaces the basic block(two $3\times3$ convolutional layers) with bottleneck block, which consists of $1\times1,3\times3$ and $1\times1$ convolution, for efficiency in the deep networks. 


Our basic CENet is built with three kinds of block: 1) initial block, 2) bottleneck block and 3) connection block. For our goal to build a small-footprint system, we choose a small channel number to avoid overfitting and heavy computational cost.
\begin{itemize}
\item\noindent \textbf{Initial block} is designed to generate the feature representations with a convolutional kernel from the MFCC feature, which includes a $3\times 3$ bias-free convolutional layer, a batch normalization layer and activation function ReLU ($\cdot$). To reduce the spatial size of the convolutional feature map, we add a $2\times2$ average pooling layer at the end of this block.

\item \noindent\textbf{Bottleneck block} is introduced to achieve the residual function with lower model complexity. For each residual function, we use a stack of 3 layers instead of a convolutional layer. The three layers are $1\times1,3\times3$ and $1\times 1$ convolutions, where $1\times 1$ layers are responsible for reducing and restoring dimensions, leaving the $3\times3$ layer a bottleneck with smaller input/output dimensions. 

\item \noindent\textbf{Connection block} is a special bottleneck block, which is used to increase dimensions and reduce the size of feature map by the convolutional layer with stride of $2$. Connection block is used at the end of each stage. 

\end{itemize}

With these three types of blocks, we are able to design several variants of the CENet at different model complexities. According to the standard ResNet architecture, our CENet is designed in a multi-stage scheme. The whole architecture consists of one initial block and three stages. Each stage includes several bottleneck blocks and one connection block. The network ends with a global average pooling layer and a $l$-way fully-connected layer with softmax function. 

We assume that the parameter size of a convolutional layer is $c_{\text{in}}\times c_{\text{out}}\times k^2$, where the $c_{\text{in}}, c_{\text{out}}$ and $k$ is the input channel number, output channel number and kernel size, respectively. Given the input feature with size of $c_{\text{in}}\times h\times w$, where the $h$ and $w$ are the height and weight, the operation for one-pass inference is $h\times w\times k^2\times c_{\text{in}}\times c_{\text{out}}$. Both the parameters and multiplication of the network are sensitive to the channel. In order to control the footprint of the network, we adopt small number of channel with different depths.


Our base model is built in the simple structure referring to the aforementioned design. Each stage comprises a few of bottlenecks and a connection block. The number of channels increases from 16 to 64 through 3 stages, which is experimentally proven efficient and effective to balance the complexity and capacity of the model. We follow equally narrow structure and explore the impact of the depth by proposing three model variants. The details of our architecture are fully depicted in Table~\ref{tab:model_details}.

\subsection{Contextual Feature Augmentation with GCN}\label{subsec:gnn}
Modeling non-local relations in feature representations is a fundamental problem in visual recognition, which enables us to capture long-range dependencies between scene entities. Recently, since the CNN-based methods are widely used in the KWS task, we explore the contextual feature augmentation to help the representations learning for KWS in this work. A promising strategy is to adopt the GCN to model the non-local relations of  the convolutional features. Below we start with a brief introduction of GCN method for estimating global context information. Then, feature augmentation strategy that combines GCN with our proposed CENet architecture is detailed.


We formulate the global contextual feature estimation~\cite{wang2018non} in the form of GCN. Formally, let $\mathbf{X}=[\mathbf{x}_1,\cdots,\mathbf{x}_N]^\top$ be a set of the convolutional features, where $\mathbf{x}_i\in \mathbb{R}^c$ is a $c$-channel feature vector and $i$ indexes the spatial location of the feature vector. In the scene of KWS, we have a conv-feature map defined on a 2D grid with size of $N \times c, N=w\times h$. 

Then, a fully-connected graph $\mathcal{G}=(\mathcal{V},\mathcal{E})$ is built with $N$ nodes $v_i\in \mathcal{V}$, and edges $(v_i, v_j)\in \mathcal{E}, \forall i<j$ (which means that this is an undirected graph), to represent the non-local relations between the features. The GCN assigns $\mathbf{x}_i$ as the input to the node $v_i$ and computes the feature representation of each node through a message passing process. The non-local relations can be defined as:

 \begin{align}
 	\tilde{\mathbf{x}}_i=\sigma\left(\frac{1}{Z_i(\mathbf{X})}\sum_{j=1}^Ng(\mathbf{x}_i,\mathbf{x}_j)\mathbf{W}^{\intercal}\mathbf{x}_j\right)\label{eq:update_onenode}
 \end{align}
 
 where $\tilde{\mathbf{x}}_i$ represents the updated feature representation at node $i$, $\sigma$ is an element-wise activation function (e.g., ReLU). $g$ is the distance measurement function encoding pair-wise relation. $Z_i(\mathbf{X})$ is a normalization factor for location $i$. $\mathbf{W}\in \mathbb{R}^{c\times c}$ is the weight matrix defining a linear mapping to encode the message from node $i$. We can write the updating equation Eq.~\ref{eq:update} in a matrix form: 
 
\begin{figure}[t]
	\begin{center}
		\includegraphics[width=0.8\linewidth]{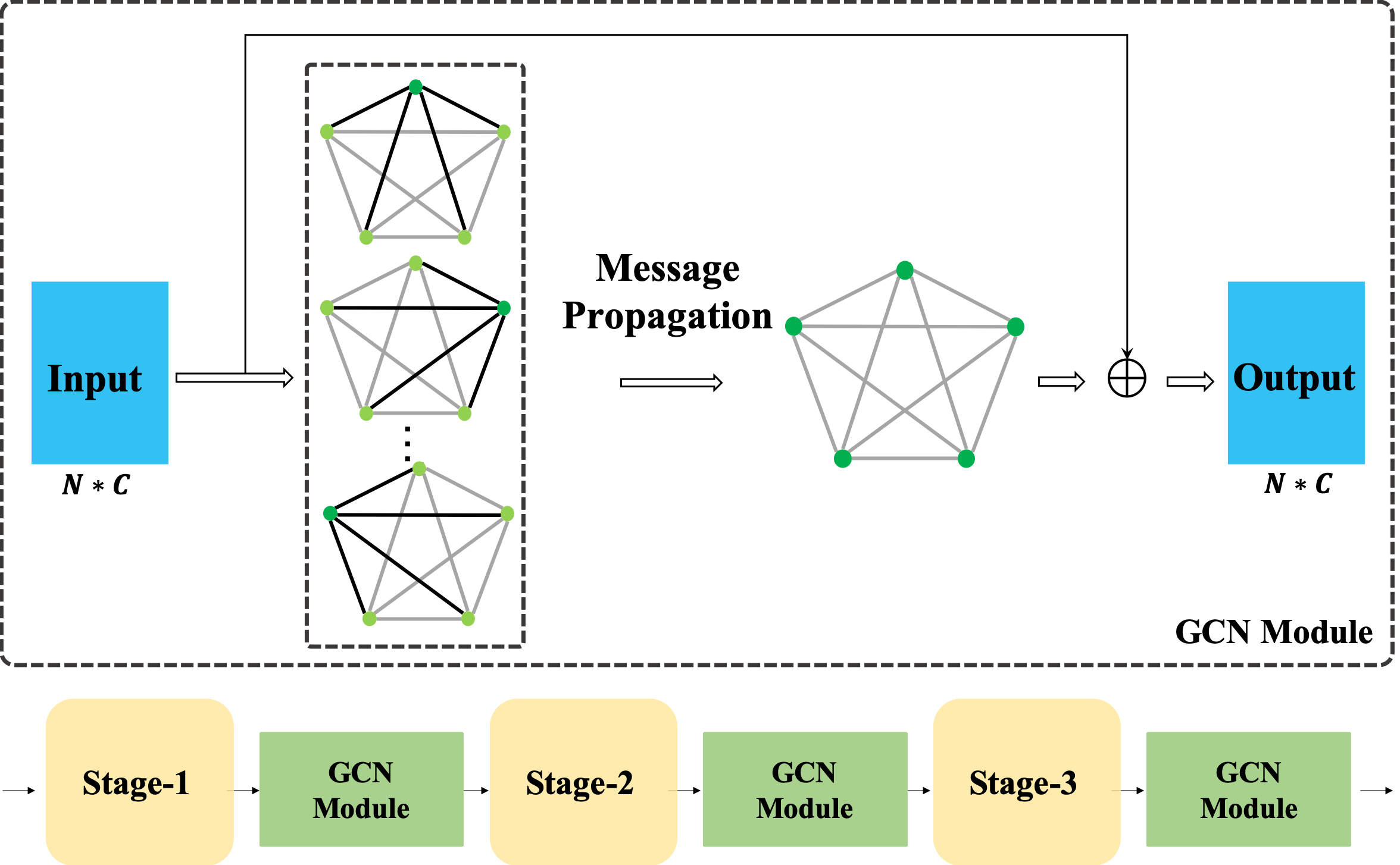}
	\end{center}
	\vspace{-2em}
	\caption{Illustration of message propagation mechanism}
	\label{fig:self_attn}
\end{figure}

\begin{align}
 	\widetilde{\mathbf{X}}=\sigma(\mathbf{A}(\mathbf{X})\mathbf{X}\mathbf{W})\label{eq:update}
 \end{align}
 
 where $\mathbf{A}(\mathbf{X})\in \mathbb{R}^{N\times N}$ denotes the affinity matrix of the graph in which $\mathbf{A}_{i, j}=\frac{1}{Z_i(\mathbf{X})}g(\mathbf{x}_i,\mathbf{x}_j)$. It is easy to extend this updating procedure to multiple iterations by unrolling the message passing into a multi-layer network~\cite{gilmer2017neural}. We focus on the single iteration setting in the remaining of this section for notational clarity. The message propagation mechanism is illustrated in the Fig.~\ref{fig:self_attn}.
 
 As the updated $\tilde{\mathbf{x}}_i$ integrates context information from entire feature set $\mathbf{X}$ through the message passing process, we can use it as an estimation of the non-local context for feature $\mathbf{x}_i$. The common strategy for defining the graph affinity matrix is based on the similarity of neighboring node features. \cite{wang2018non} proposes multiple choices of pairwise function $g$:
 	\begin{itemize}
 		\item\noindent  \textbf{Gaussian}:
 		\begin{align}
 			g(\mathbf{x}_i, \mathbf{x}_j)=e^{\mathbf{x}_i^{\intercal}\mathbf{x}_j}
  		\end{align}
  	    here $\mathbf{x}_i^{\intercal}\mathbf{x}_j$ is dot-product similarity and the normalization factor is $Z_i(\mathbf{X})=\sum_{\forall j}g(\mathbf{x}_i, \mathbf{x}_j)$.

 		\item \noindent \textbf{Embedded Gaussian}: 
 		\begin{align}
 			g(\mathbf{x}_i, \mathbf{x}_j)=e^{\theta(\mathbf{x}_i)^{\intercal}\phi(\mathbf{x}_j)}
 		\end{align}
 	
 	where $\theta(\mathbf{x}_i)=\mathbf{W}_\theta \mathbf{x}_i$ and $\phi(\mathbf{x}_j)=\mathbf{W}_\phi\mathbf{x}_j$ are two embeddings. $Z_i(\mathbf{X})=\sum_{\forall j}g(\mathbf{x}_i, \mathbf{x}_j)$. 

 	\end{itemize}
 	Specifically, we adopt the embedded Gaussian with softmax function to measure the pairwise similarity. Thus the Eq.~\ref{eq:update} can be rewritten as:
 	  \begin{align}
 	  \widetilde{\mathbf{X}} = \sigma(\text{softmax}(\mathbf{X}^{\intercal}\mathbf{W}_\theta^{\intercal}\mathbf{W}_{\phi}\mathbf{X})\mathbf{X}\mathbf{W})
 	  \end{align}


 This feature augmentation strategy considers all positions into the operation for modeling the global context, which is different from a fully connected (fc) layer. Given the context feature $\widetilde{\mathbf{X}}$, we adopt a simple feature augmentation strategy to integrate the contextual representation with the original convolutional feature. We use a weighted summation of the two features as in~\cite{fu2019dual}:
\vspace{-1mm}
\begin{align}
	\mathbf{X}_a = \gamma \widetilde{\mathbf{X}} + \mathbf{X}
\end{align} 
 \vspace{-1mm}

where $\gamma$ is a scaling parameter to be learned and $\mathbf{X}_a$ is the augmented feature. 


 We can easily insert the GCN module into our CENet. In practical, we insert the module at the end of each stage to encode the long range dependencies in different levels. Specifically, we use only one-layer GCN to incorporate the non-local relations while maintaining small model complexity.

\section{Experiment}\label{sec:exp}
In this section, we conduct a series of experiments on the Google Command Dataset~\cite{warden2018speech} to validate the effectiveness of our method.
Firstly, we present the introduction of the dataset and the implementation details in \textbf{Sec.~\ref{subsec:detail}}. 
Quantitative results and ablation study are given in \textbf{Sec.~\ref{subsec:results}}. 
In \textbf{Sec.~\ref{subsec:visualization}}, the augmented convolutional feature map is visualized to study the effectiveness of GCN module. We also plot receiver operating characteristic (ROC) curves for comprehensive analysis.


\subsection{Experimental Configuration}\label{subsec:detail}
We use the Google Speech Command Dataset to evaluate our method. The dataset consists of 65000 one-second fixed length utterances, which concludes 30 short words from thousands of people. Following the experimental setting of~\cite{zhang2017hello, tang2018deep}. We focus on discriminating 12 commands: "yes", "no", "up", "down", "left", "right", "on", "off", "stop", "go", unknown, or silence. The dataset is  split into 80:10:10 for training, test and validation by the SHA1-hashed name of
audio files. Thus there are no overlapping speakers between the train, test and validation sets. We randomly add the background noise in the dataset to the training audios with a probability of 0.8, which is intended to enhance the data. The SNR is randomly selected between $\left[ 5,15 \right]$ db. A random $Y$ ms time shift is implemented before transforming the audio to MFCCs, where $Y \sim \text{Uniform} [-100,100]$.

For the model architecture, our system is implemented with PyTorch auto-differentiable framework. There are three variants of the CENet models. Then, we insert the GCN module at the end of each stage. 
The learnable parameters of our model are randomly initialized. Our models are trained with SGD optimizer, and the total training epochs is 350. We use the "poly" learning rate policy where current learning rate equals to the base one multiplying $(1-\frac{iter}{maxiter})^{power}$. We set the base learning rate to 0.01 and power to 0.9. The batch size is 64 and $L_2$ weight decay is $10^{-3}$.

\subsection{Quantitative Results}\label{subsec:results}
We report the quantitative results for the aforementioned models in Table~\ref{tab:quant_results}. Following the evaluation method in~\cite{tang2018deep}, we use the class accuracy as the evaluation metric, which is measured as the fraction of the correct prediction. We also compare parameter numbers and multiplications with other baseline methods.

\begin{table}[t]
	\center
	\caption{Performance of CENet and Baseline Method}
	\vspace{2mm}
	\resizebox{0.4\textwidth}{!}{
		\begin{tabular}{l|ccc}
			\toprule[0.3mm]
			Model & \#Param. & Mult. & Results \\
			\midrule
			trad-fpool13\cite{sainath2015convolutional} & 1.37M & 125M & 90.5\% \\
			tpool2\cite{sainath2015convolutional} & 1.09M& 103M & 91.7\% \\
			one-stride1\cite{sainath2015convolutional} & 954K& 5.76M & 77.9\% \\
			\midrule
			res15\cite{tang2018deep} & 238K & 894M & 95.8\% \\
			res8\cite{tang2018deep} & 110K & 30M & 94.1\% \\
			res15-narrow\cite{tang2018deep} & 42.6K & 160M & 94.0\% \\
			res8-narrow\cite{tang2018deep} & 19.9K & 5.65M & 90.1\% \\
			\midrule
			DS-CNN-S\cite{zhang2017hello} & 38.6K & 5.4M & 94.4\% \\
			DS-CNN-M\cite{zhang2017hello} & 189.2K & 19.8M & 94.9\% \\
			DS-CNN-L\cite{zhang2017hello} & 497.6K & 56.9M & 95.4\% \\
			\midrule 
			CENet-6  & \textbf{16.2K} & \textbf{1.95M} & 93.9\% \\
			CENet-24  & 44.3K & 8.51M &  95.6\%  \\
			CENet-40  & 61K & 16.18M & \textbf{96.4\%} \\
			\bottomrule[0.3mm]
		\end{tabular}
		\label{tab:quant_results}}
\end{table}

		\vspace{-1em}
\subsubsection{Comparison with Prior Works} 
		\vspace{-0.5em}

We first investigate different CENet variants compared with the recent works. The performance indicates our CENet can outperform the state-of-the-art methods with fewer parameters and less computations.

Concretely, our base model $\text{CENet-24}$ can achieve the comparable performance with res15~\cite{tang2018deep}. Compared with res15~\cite{tang2018deep}, our model achieves a 5$\times$ reduction in model complexity with only 44k parameters and the $>$100$\times$ reduction multiplies in the feedforward inference pass. This result indicates the effectiveness of bottleneck architecture in building efficient network.

Moreover, we also investigate the impact of the depth by changing the number of blocks in different stages. We firstly reduce the number of blocks in each stage by 4 times and get the $\text{CENet-6}$, which is the most compact network in this study. Compared with other compact models, like $\text{one-stide1}$  in~\cite{sainath2015convolutional} and $\text{res8-narrow}$ in~\cite{tang2018deep}. CENet-6 can achieve 93.9\% in accuracy, which outperforms the other two models by a large margin with only 16.2K parameters. 

Finally, we stack more layers to build a more powerful variant. It is worth noting that the model complexity is sensitive to the number of blocks of last stage due to the large channel number. We only increase the number of blocks in the first two stages to get CENet-40, which could achieve a higher performance with 96.4\%. Despite our changes, CENet-40 still maintains a small number of parameters (61K) and operations (16.18M). All three models demonstrate the effectiveness and efficiency of our CENet.

		\vspace{-1em}
\subsubsection{Results of the CENet with GCN Module}
		\vspace{-0.5em}

\begin{table}[t]
	\small
	\center
	\caption{Test Accuracy and footprint of different models}
	\resizebox{0.35\textwidth}{!}{
		\begin{tabular}{l|ccc}
			\toprule[0.3mm]
			Model & \#Param. & Mult. & Results \\ 
			
			\midrule
			CENet-6   & \textbf{16.2k} & \textbf{1.95M} & 93.9\% (94.9\%)\\
			CENet-GCN-6  & 27.6k & 2.55M & 95.2\% (95.7\%)\\
			\midrule
			CENet-24   & 44.3k & 8.51M & 95.6\% (96.1\%)\\
			CENet-GCN-24  & 55.6k & 9.11M & 96.5\% (96.5\%)\\
			\midrule
			CENet-40   & 60.9K & 16.18M & 96.4\% (96.7\%)\\
			CENet-GCN-40  & 72.3K & 16.78M & \textbf{96.8}\% (\textbf{97.0}\%)\\ 
			\bottomrule[0.5mm]
		\end{tabular}
		\label{tab:addGCNresult}}
\end{table}

To validate the contextual feature augmentation mechanism, we evaluate the performance of CENet-GCN models. We insert the GCN module at the end of each stage to get CENet-GCN-6, CENet-GCN-24 and CENet-GCN-40, respectively. The CENet-GCN models are trained in the same way as CENet without additional supervision, and the results are shown in Table~\ref{tab:addGCNresult}. 

Interestingly, the CENet-6 can achieve a 1.3\% improvement with GCN module with 10K additional parameters, which uses only 27.6K parameters to get the comparable accuracy with the res15 and CENet-24. The CENet-GCN-24 is able to outperform the CENet-40 with a more compact model and less computation. 

Compared with CENet models, CENet-GCN-24 and CENet-GCN-40 can also achieve a 0.9\% and 0.4\% improvement, respectively. Due to a larger capacity for the contextual feature learning with a larger effective receptive field, the performance gain decreases in the deeper network. The results of three model variants demonstrate that the GCN module is capable of incorporating the contextual information to improve the feature learning with few parameters and small computation cost.

To compare our method with other state-of-the-art approaches~\cite{tang2018deep} under the same setting, which uses the MFCC feature, thus we also adopt the MFCC feature as the input in all our experiments. We also evaluate the impact of input feature. MFCC feature is de-correlated in the frequency domain when extracted, and fbank feature still remains the temporal-frequency correlation of spectral representations.  In principle, the convolutional layers can learn better from fbank feature because it actually can leverage the spatial-temporal correlations of spectral representations. We further demonstrate our method with the fbank feature as the input and report the results in Table~\ref{tab:addGCNresult}, which are shown inside the brackets. Compared with the results of the MFCC feature, the quantitative performance indicates our approaches will achieve 0\%-1.0\% improvement in all model variants with fbank feature as the input. 
\begin{table}[t]
	\small
	\center
	\caption{Results of the CENet with GCN in different stage.}
	\resizebox{0.4\textwidth}{!}{
		\begin{tabular}{l|cccc}
			\toprule[0.3mm]
			Model & Stage &\#Param. & Mult. & Results \\ 
			
			\midrule
			CENet-6 & - & \textbf{16.2k} & \textbf{1.95M} & 93.9\% \\
			
			CENet-GCN-6 & Stage-1  & 17.8K & 2.27M & 94.3\% \\
			
			CENet-GCN-6 & Stage-2   & 19.8K & 2.13M & 95.0\% \\
			
			CENet-GCN-6 & Stage-3  & 22.5K & 2.05M & 94.4\% \\
			
			CENet-GCN-6 & Stage-1,2,3  & 27.6K & 2.55M & \textbf{95.2}\% \\
			\bottomrule[0.5mm]
		\end{tabular}
		\label{tab:ablation_study}}
\end{table}

		\vspace{-1em}
\subsubsection{Different Stage with GCN Module}
		\vspace{-0.5em}

Furthermore, we investigate the stage which we should add the GCN module to augment the feature with contextual information. We insert one GCN module for different residual stages, based on the CENet-6 backbone, to build the network. The quantitative results which shown in Table~\ref{tab:ablation_study} demonstrate the effects of the proposed GCN for each of stage and all stages.

Noted that incorporating GCN module into the stage-1 gain smaller compared with the base method, we can analyse the lower-level feature maps lack semantic information. Meanwhile, the stage-2 which equipped with a single GCN module could achieve 1.1 points improvement for accuracy. Moreover, it is straightforward to incorporate our GCN module into multiple residual stages in order to augment the feature maps with multi-level contextual information. Our results show that adding GCN module to all stages (1,2 and 3) together can achieve 95.2 for accuracy, which obtains 1.3 performance gain, respectively. The quantitative results indicate that our GCN can achieve a large margin improvement with less computational cost. 

\begin{figure}[t]
	\begin{center}
		\includegraphics[width=0.75\linewidth]{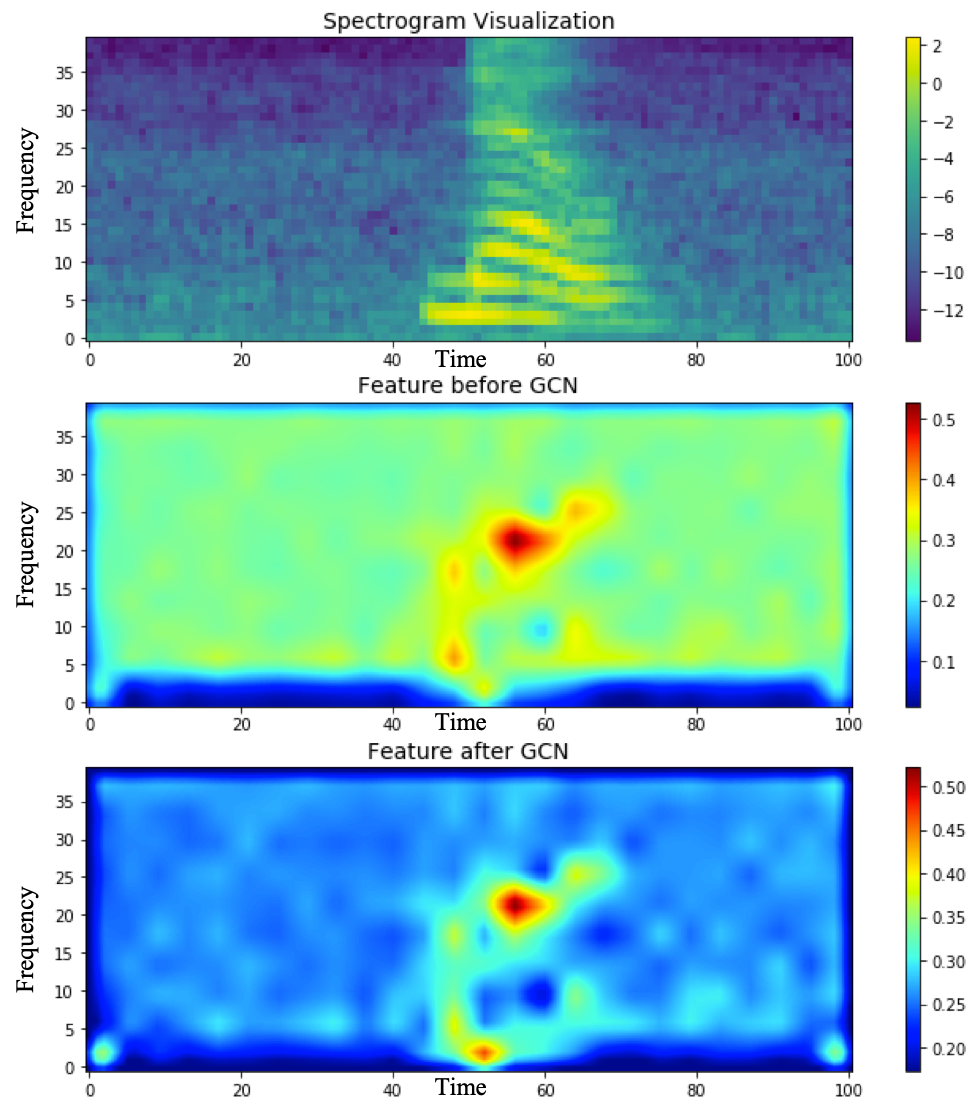}
	\end{center}
		\vspace{-2em}
	\caption{Visualization of the Feature Map. The first row is the spectrogram visualization of the audio sequence. The second row is the averaged feature map of the stage-1 before the GCN. The last row is the averaged feature map of the stage-1 after the GCN. Feature maps are generated from trained CENet-GCN-6 model.}
	\label{fig:feature_vis}
\end{figure}

\vspace{-1em}

\subsection{Visualization and Analysis}\label{subsec:visualization}

		\vspace{-0.5em}
\subsubsection{Visualization of Feature Map }
		\vspace{-0.5em}

We visualize the convolutional feature map in Fig.~\ref{fig:feature_vis} for better understanding the contextual feature augmentation mechanism proposed in our work. The convolutional feature maps are averaged along the channel dimension, and interpolated with same size of the mel-scale spectrogram for visualization. Compared with the feature map before GCN, it is evident that the GCN module utilizing the contextual information can help highlight the most discriminative region and enlarge the gap between the voiced and non-voiced regions.

		\vspace{-1em}
\subsubsection{ROC Curve Analysis}
		\vspace{-0.5em}

\begin{figure}[t]
	\begin{center}
		\includegraphics[width=0.75\linewidth]{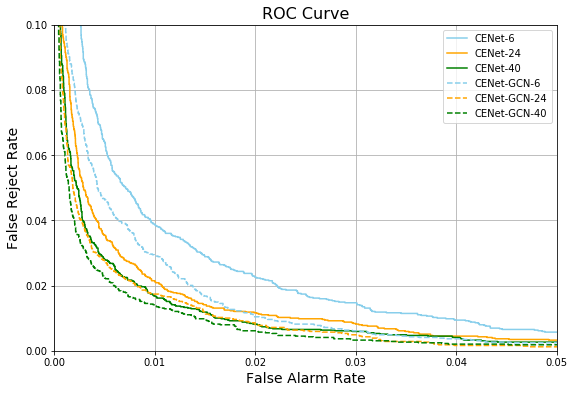}
	\end{center}
		\vspace{-2em}

	\caption{ROC curve of CENet and CENet-GCN models.}
	\label{fig:GCN_roc}
\end{figure}

Furthermore, we plot the receiver operating curve (ROC) for extensive analysis. The $x$ axis is false alarm rate (FAR) and the $y$ axis is false reject rate (FRR), representing the probability of false positives and the probability of false negatives, respectively.
For a given sensitivity threshold—defined as the minimum probability which a class is considered positive during evaluation. Curves for each of the keywords are computed by sweeping the sensitivity threshold [0.0, 1.0] and then being averaged vertically to produce the overall curve for a particular model. The model with less area under the curve (AUC) are the better. Curves of all variants of our CENet and CENet-GCN are plotted in Fig.~\ref{fig:GCN_roc}. The ROC curves demonstrate the effectiveness of the GCN module, which is consistent with the quantitative results in the Table~\ref{tab:addGCNresult}.

\begin{figure}[t]
	\begin{center}
		\includegraphics[width=0.75\linewidth]{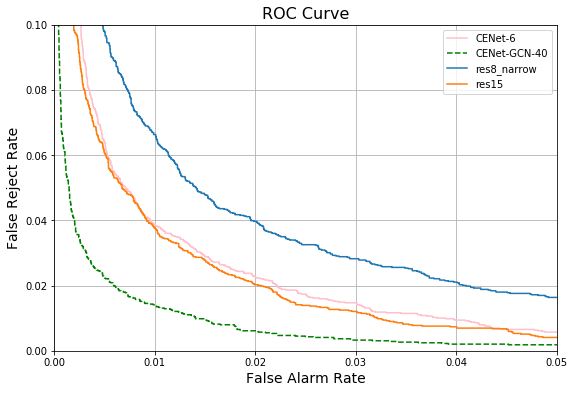}
	\end{center}
		\vspace{-2em}
	\caption{ROC curve for different models}
	\label{fig:comparition_roc}
\end{figure}

We compare our most compact model and most powerful model with prior works, as shown in Fig.~\ref{fig:comparition_roc}. Both of our models can outperform baseline methods with a large margin, respectively.

\vspace{-3mm}
\section{Conclusion}\label{sec:con}
We introduce a novel efficient network for KWS tasks, which leverages the power of residual connection with the bottleneck structure and graph convolutional network method. Based on our proposed network structure, we build several variants of our proposed model with different model complexity. Our models are evaluated on the Google Speech Command Dataset. Our basic CENet models outperform the current state-of-the-art method with fewer parameters and simpler network structure. To get further, we introduce the graph convolutional network to KWS task to encode context of features. Models combined with graph convolutional network method can perform even better than our basic models. 


%
%
\newpage
\bibliographystyle{IEEEbib}

\bibliography{mybib}

\end{document}